\documentclass[aps,prd,floatfix,superscriptaddress,twocolumn,showpacs,amssymb]{revtex4-1}
\usepackage{amssymb,amsmath,verbatim}
\usepackage{graphicx}
\usepackage{epsfig}
\usepackage{dcolumn}
\usepackage{bm}

\newcommand{\be}{\begin{equation}}
\newcommand{\ee}{\end{equation}\noindent}
\newcommand{\eei}{\end{equation}}
\newcommand{\bea}{\begin{eqnarray}}
\newcommand{\eea}{\end{eqnarray}\noindent}
\newcommand{\eeai}{\end{eqnarray}}

\newcommand{\hf} {\frac{1}{2}}
\newcommand{\nn}{\nonumber\\}

\def\eq#1{(\ref{#1})}

\def\t#1{{\tilde#1}}
\def\c#1{{\cal#1}}
\def\b#1{{\bar#1}}

\def\u#1{{\underline{#1}}}

\begin{document}
\title{Triple point in the $O(2)$ ghost model with higher-order gradient term
}

\author{Z. P\'eli}
\affiliation{Department of Theoretical Physics, University of Debrecen,
P.O. Box 5, H-4010 Debrecen, Hungary}

\author{S. Nagy}
\affiliation{Department of Theoretical Physics, University of Debrecen,
P.O. Box 5, H-4010 Debrecen, Hungary}
\author{K. Sailer}
\affiliation{Department of Theoretical Physics, University of Debrecen,
P.O. Box 5, H-4010 Debrecen, Hungary}

\date{\today}

\begin{abstract}
The phase structure and the infrared behaviour of the Euclidean 3-dimensional $O(2)$ symmetric ghost scalar field $\phi$  has been investigated in Wegner and Houghton's renormalization group framework,  including higher-derivatives in the kinetic term. It is pointed out that higher-derivative coupling provides three phases and leads to a triple point in that RG scheme. The types of the phase transitions have also been identified.

\end{abstract}



\maketitle

\section{Introduction}

It is well-known that the existence of the triple point, the point of coexistence of three phases is very common in condensed matter physics, generally realized as the coexistence of the gaseous, liquid and solid phases of the same material, but also occuring
in magnetic materials with more than one solid phases in equilibrium \cite{ChaLub2000,Chat2006} as well as in $^4$He as the equilibrium of two solid and a liquid or that of the supefluid, normal fluid and solid phases \cite{Lebel1992}. Now we shall show that in a particular approximation of the functional renormalization group 
approach one finds that the 3-dimensional Euclidean $O(2)$ symmetric ghost scalar model with wavefunction renormalization $Z=-1$ and the higher-derivative term  $Y\phi \Box^2 \phi$ exhibits a triple point where
the symmetric phase, the symmetry broken phase, and the phase with restored symmetry coexist in equilibrium. In general, field theory models with higher-derivative terms of alternating signs have rather rich phase structure corresponding to  various periodic structures \cite{Fin1997,Bran1999,Four2000,Basar2009}. Existence of the triple point in ordinary $O(2)$ symmetric models with appropriate higher-derivative terms has also been shown in \cite{Pol2005}.

The phase structure of the ghost $O(2)$ model has been analysed by us in the framework of 
Wegner and Houghton's (WH) renormalization group (RG) method \cite{Weg1973} with the sharp gliding momentum cutoff $k$. Using WH RG framework one is restricted
  to the local potential approximation (LPA), the lowest order of the gradient expansion. In the LPA the wavefunction renormalization $Z$ and couplings of the higher derivative terms do not acquire any RG flow. Since the wavefunction renormalization $Z$ is dimensionless, it can be kept constant unambiguously like $Z=+1$ for ordinary and $Z=-1$ for ghost models. There occurs, however, an ambiguity when the couplings of higher-derivative terms are accounted for which have nonvanishing momentum dimensions. It corresponds to different approximations or RG schemes to keep either the dimensionful, or the dimensionless higher-derivative couplings constant. In our previous paper \cite{Peli2016} we argued for keeping the dimensionful coupling $Y$ constant and showed that the model exhibits two phases: besides the trivial symmetric phase there occurs a phase with restored symmetry characterized by a quasi-universal dimensionful effective potential. The existence of the latter is related with the occurrence of the ghost condensate at intermediate scales $k$.
Now we shall take another point of view and keep the {\em dimensionless} coupling  $\t{Y}=Yk^{-2}$  constant during the WH RG flow.
 In this approximation it shall be shown that the model  has three phases and exhibits the possibility of the coexistence of all three phases in equilibrium.

\section{Wegner-Houghton renormalization group for the ghost $O(2)$ model}

In this paper we study of the 3-dimensional, Euclidean, $O(2)$ symmetric model for the real two-component ghost scalar field $\underline{\phi}
=\begin{pmatrix}\phi_1\cr\phi_2\end{pmatrix}$ using the ansatz
\bea \label{realaction}
S_k[\underline{\phi}]& =& \frac{1}{2}\int d^3 x \underline{\phi}^T
\Omega(-\Box )\underline{\phi} + \int d^3 x U_k( \u{\phi}^T\u{\phi} ),
\eea
for the blocked action in LPA,
where $U_k( \u{\phi}^T\u{\phi} )$ stands for the blocked potential
 assumed to be of the  polynomial form (a Taylor expansion truncated at the order $\phi^{2M}$)
\bea\label{polpot}
 U_k(r)&=&\sum_{n=0}^M\frac{v_n(k)}{n!} r^{n}
\eea
with  $r=\hf \u{\phi}^T\u{\phi}$ and
\bea 
\Omega(-\Box )&=&-Z\Box+Y\Box^2
\eea
with the wavefunction renormalization $Z=-1$ and the higher-derivative coupling
$Y>0$. The phase structure of the model has been analysed in the framework of 
WH RG method with the sharp gliding momentum cutoff $k$. In the LPA the wavefunction renormalization $Z$ and 
the higher-derivative coupling $Y$ do not acquire RG flow. The WH RG equation
for the local potential is given as \cite{Peli2016}
\bea\label{WHRGeq}
  k\partial_k U_k(r) &=& -\alpha k^3 \lbrack \ln s_+(k)+ \ln s_-(k) \rbrack,
\eea
where
\bea
 s_+(k)&=& \Omega( k^2) +U^\prime_k(r) + 2rU^{\prime\prime}_k(r),\nn
 s_-(k)&=& \Omega( k^2) +U^\prime_k(r)
\eea
with $U^\prime_k(r)=\partial_rU_k(r)$, $U^{\prime\prime}=\partial_r^2 U_k(r)$, and
$\alpha= 1/(4\pi^2)$.
Here $r$ corresponds now to a constant background field  $\Phi$ with $\Phi=\sqrt{2r}\ge 0$ pointing in an arbitrary direction $\u{e}$ in  the internal space. That background field is the tool to find out the form of the potential.

 In general,
 the WH RG equation may lose its validity. This happens when at least one of the arguments of the logarithms in the right-hand side of Eq. \eq{WHRGeq} seases to be positive at some nonvanishing scale $k_c$. For scales $k\le k_c$ the resummation of the loop expansion by means of the WH equation is not any more possible. The IR behaviour can then be 
revealed by means of the tree-level renormalization (TLR) procedure  \cite{Ale1999} (see also its application to the $O(2)$ model in our previous paper \cite{Peli2016}). While in the case of the one-component $(N=1)$ real scalar field the vanishing of $s_+(k)$ governs the singularity, in cases with $N\ge 2$ the vanishing of $s_-(k)$ does it. The critical scale $k_c$ is given by $s_-(k_c)|_{\Phi=0}=0$ implying $Z+\t{Y}+\t{v}_1(k_c)=0$, just like in the case  $N=1$.
(Throughout this paper the dimensionless quantities shall be denoted by tilde, so that $\phi= k^{1/2}\t{\phi}$, $v_n=k^{3-n}\t{v}_n$, $U_k=k^3\t{U}_k$, and $Y=k^{-2}\t{Y}$.) The spinodal instability  at the singularity scale $k_c$ 
reveals itself in building up an inhomogeneous field configuration $\underline{\psi}$ on the 
homogeneous  background. The essence of TLR is to decrease the scale $k$ by a
step $\Delta k\ll k$ and to find out the inhomogeneous configuration $\underline{\psi}$ that minimizes the Euclidean action at the given scale $k<k_c$, that determines the blocked action at the lower scale $k-\Delta k$ via
\bea\label{tlrbl}
  S_{k-\Delta k} [\underline{e}\Phi]&=&{\rm{min}}_{\underline{\psi}} S_k[\underline{e}\Phi+\underline{\psi}] .
\eea 
We shall restrict the function space of spinodal instabilities to those of stationary waves $\underline{\psi}$ pointing into the direction $\underline{e}$ of the homogeneous background field in the internal space and describing sinusoidal periodicity  in a given direction $n_\mu$  of the external space, i.e., to the form
\bea\label{tlransatz}
 \underline{\psi} &=&  \underline{e}2\rho \sin( k n_\mu x_\mu  +\theta)
\eea
with the phase shift $\theta$.  Making use of the ansatz \eq{tlransatz}
the TLR blocking relation \eq{tlrbl} reduces to the recursion relation
\bea\label{tlrpotrec}
U_{k-\Delta k}(\Phi)&=&\min_{\{\rho\}}\biggl(U_k(\Phi)+(Z+ \t{Y})k^2\rho^2\nn
&&+\sum_{n=1}^M \frac{\rho^{2n}}{(n!)^2}\partial_\Phi^{2n} U_k(\Phi)\biggr).
\eea 
The ansatz \eq{tlransatz} reduces the TLR of the $O(2)$ model to that of the 
$O(1)$ model with the same wavefunction renormalization $Z$ and  higher-derivative coupling $\t{Y}$.
Let $\rho_k(\Phi)$ be the amplitude minimizing the expression in  the right-hand side of the recursion relation \eq{tlransatz}. Clearly, a nonvanishing value of $\rho_k(\Phi)$ breaks $O(2)$ symmetry in internal space, as well as $O(3)$  and translational symmetries in the 3-dimensional external space. For local potentials of the form \eq{polpot}  and for scales $k<k_c$ the interval $0\le \Phi\le \Phi_c(k)$ (with $\Phi_c(k)=\sqrt{k}\t{\Phi}_c(k)$), in which the  instability 
occurs, is determined via the relation $s_-(k)=0$  as
\bea\label{phico2}
  \t{\Phi}_c(k) &=& 
 \sqrt{ - \frac{ 2  \lbrack Z+\t{Y} +\t{v}_1(k) \rbrack }{
 3 \t{v}_2(k) } } .
\eea

Our numerical procedure for the determination of the RG trajectories 
is just the same as in our paper \cite{Peli2016}. The  WH RG equations
are rewritten as a coupled set of ordinary differential equations for the
dimensionless couplings $\t{v}_n$ of the local potential $\t{U}_k(\t{\Phi})$
and those solved with the truncation $M=10$ with fourth-order Runge-Kutta method  for scales $k_c<k\le \Lambda$. It may happen that it holds the inequality
 $s_-(\Lambda)<0$ at the ultraviolet (UV) cutoff scale $\Lambda$ already. Therefore, we define the singularity scale as $k_s=k_c$ for the cases with $s_-(\Lambda)>0$ and $k_s=\Lambda$ for cases with  $s_-(\Lambda)<0$. If there is a singularity
then TLR is applied for scales $k<k_s$ in order to determine the RG flow in the IR regime by means of the recursion relation \eq{tlrpotrec} rewritten in terms of the dimensionless quantities. The scale $k$ has then been decreased from the scale $k_s$  by at least two orders of magnitude with the  step size $\Delta k /k = 0.005$ and the truncation $M=10$.    The numerical precision was set to $80$ digits. Generally $\sim 1000$ iteration  steps have been numerically performed at each value of the constant background $\Phi$ for the minimization of the blocked potential $U_k(\rho, \Phi)$ with respect to the amplitude $\rho$ of the spinodal instability.  The  minimization with respect to $\rho$ in the right-hand side of Eq. \eq{tlrpotrec} and the determination of the couplings at the lower scale $k-\Delta k$ with least-square fit  are performed in the interval  $0\le \Phi\le\b{\Phi}$ of the background fields which has been chosen in a similar manner as described in \cite{Peli2016}. Namely,  for `Mexican hat'
like potential $U_{k_s}(\Phi)$  the choice $\b{\Phi}\approx 1.5 \Phi_m$ has been made where $\pm\Phi_m$ are the  positions of the local minima of the potential with $\Phi_m= \sqrt{ -2v_1(k_s)/3v_2(k_s)} $. For convex potentials $U_{k_s}(\Phi)$
the choice $\b{\Phi}\gtrsim 30$ has been made. It has been observed numerically that the blocked potential does not acquire tree-level corrections outside of the interval $0\le\Phi\le \Phi_c$  with $\Phi_c$ given by Eq. \eq{phico2}, but the choice of the larger interval makes the minimization and fitting
numerically stable.

\section{Phase structure and IR scaling laws}

\subsection{Phase diagram}

\begin{figure}[htb]
\centerline{
\psfig{file=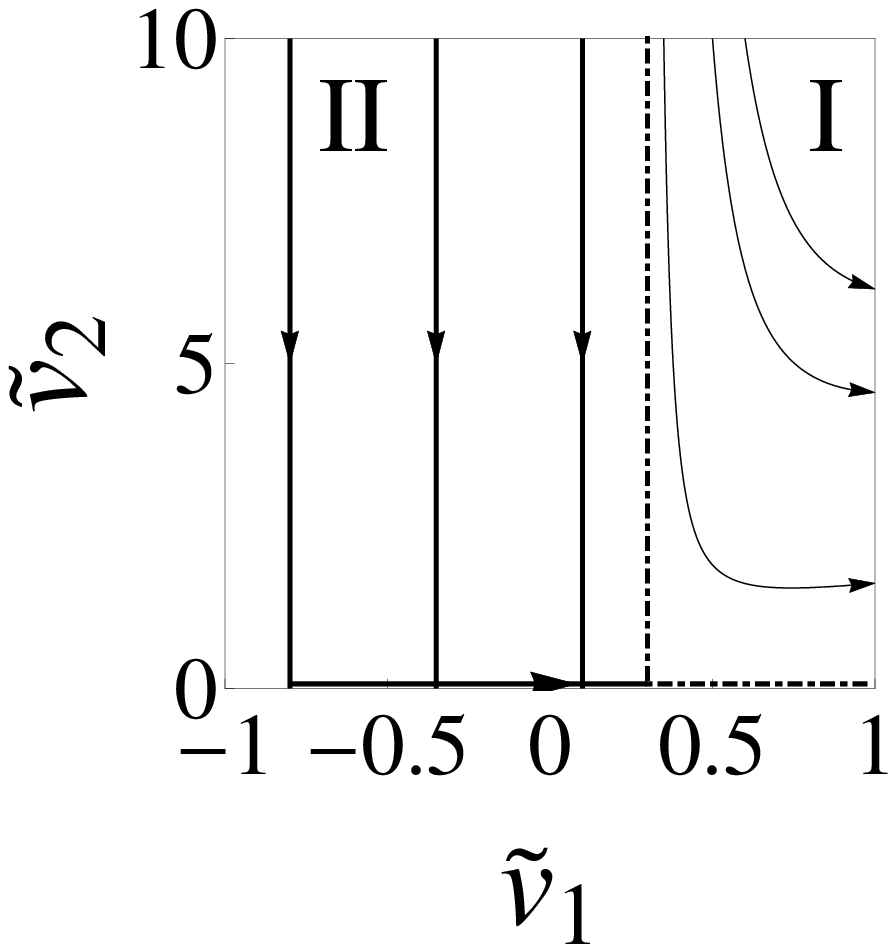,height=3.52cm,width=4.4cm,angle=0}
\psfig{file=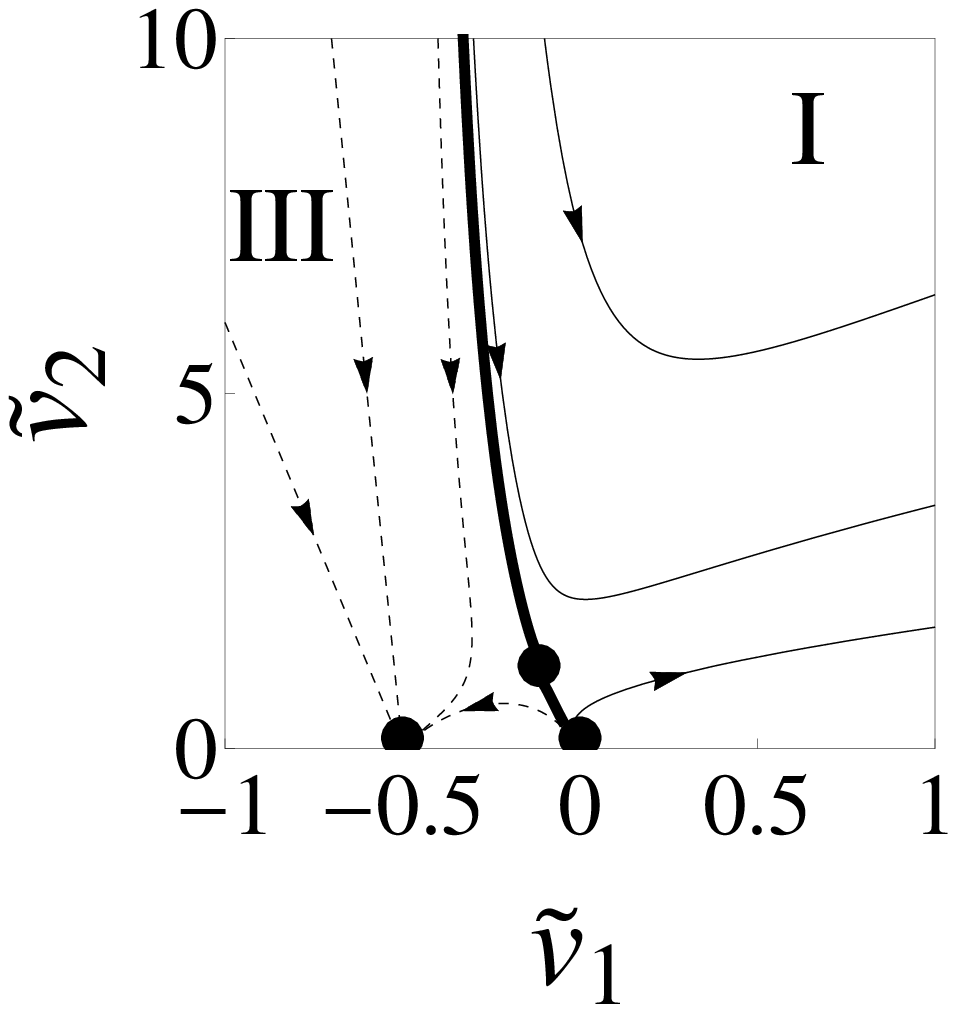,height=3.52cm,width=4.4cm,angle=0}
}
\centerline{
\psfig{file=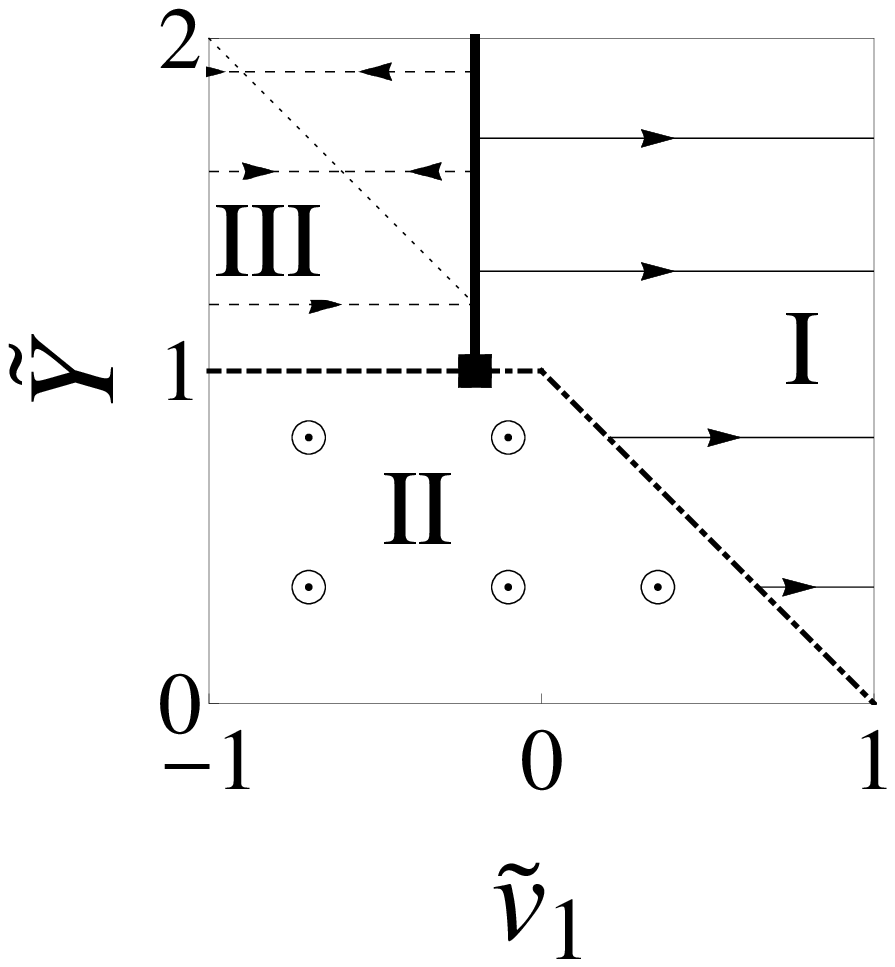,height=3.52cm,width=4.4cm,angle=0}
\psfig{file=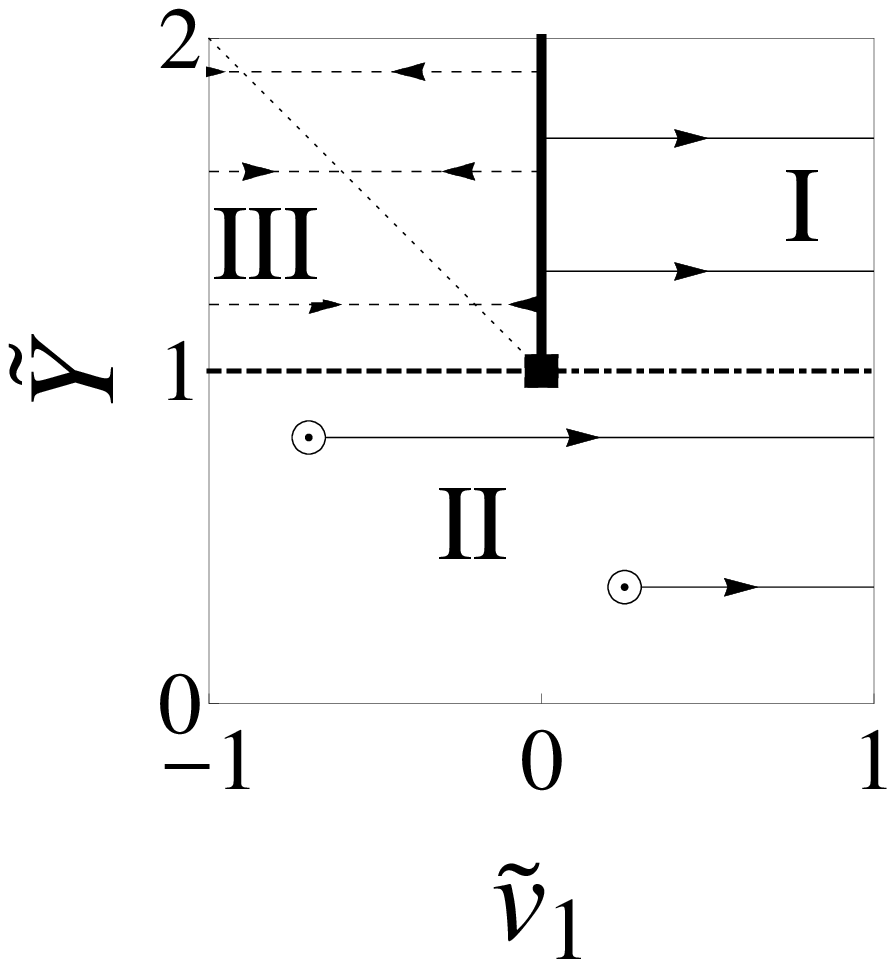,height=3.52cm,width=4.4cm,angle=0}
}
\caption{\label{fig:phase_structure}  Various planar slices of the phase diagram of the ghost $O(2)$ model with a few typical RG trajectories in the parameter space $(\t{v}_1, \t{v}_2,\t{Y})$:
the slice $(\t{v}_1, \t{v}_2)$ for $\t{Y}=0.7$ (at the top to the left), the slice  $(\t{v}_1, \t{v}_2)$ for  $\t{Y}=1.5$ (at the top to the right) with the fixed points (dots),
 the slice $(\t{v}_1, \t{Y})$ for $\t{v}_2>0$  (at the bottom to the  left),
and the slice $(\t{v}_1, \t{Y})$ for $\t{v}_2=0$ (at the bottom to the right).
The phase boundaries II- I, III-I and III-II are depicted by thick dashed-dotted, thick full, and  dashed lines, respectively. The  dotted line
 represents a section of the straight line  $\t{v}_1 =1-\t{Y}$, which is the IR fixed line in the slice with $\t{v}_2=0$; the full square
stands for the triple point. The dotted circles represent RG trajectories running perpendicularly to the $\t{v}_2=$const. planes.
}
\end{figure}
The phase structure has been investigated for RG trajectories started
in the hypercube $ [-1,+1]\otimes [0,10]\otimes [0,2]$
 in the 3-dimensional parameter space $(\t{v}_1, \t{v}_2, \t{Y})$. Various
 slices of the phase diagram 
are shown in Fig. \ref{fig:phase_structure}.  The  identification of the phases is based on the concept of the so-called sensitivity matrix \cite{Pol2003,Nagysens}. The matrix $S_{n,m}$ is built up by the derivatives of the running coupling constants with respect to  the bare ones
\be
S_{n,m} = \frac{\partial g_n(k)}{\partial g_m(\Lambda)}.
\ee
We can find different phases when a singularity takes place in the IR ($k\to 0$) and the UV ($\Lambda\to\infty$) limits of the elements of the sensitivity matrix. According to such type of identification we can find different phases in the model when the effective potential depends on different sets of  bare couplings.
 Using this technique we found that
there exist three phases and a triple point in all slices at constant $\t{v}_2$.  It shall be shown that there is a symmetric phase (phase I), a phase with restored symmetry (phase II), and a phase with spontaneously broken symmetry (phase III).
Some remarks should be made with respect to the phase diagram. In our WH RG approach all RG trajectories lie in one of the $\t{Y}=$const. planes.
The RG trajectories belonging to phase II  arrive perpendicularly to the
plane $\t{v}_2=0$, where they make a turn  with 90$^o$ and run away to plus infinity parallel to the $\t{v}_1$ axis. This happens because the dimensionful coupling $v_1$ takes a nonvanishing constant value in the IR limit $k\to 0$. Therefore the phase boundary II-I is the 2-dimensional surface $\bigl(\t{v}_1=1-\t{Y}, ~\t{v}_2>0,~ 0<\t{Y}<1\bigr)\cup \bigl( 1-\t{Y}<\t{v}_1\le 1,~\t{v}_2=0, ~ 0<\t{Y}<1\bigr) $. The Gaussian and Wilson-Fisher fixed points shown in the top-right subfigure in  Fig. \ref{fig:phase_structure} belong to phase I and stand for fixed lines with any values of $\t{Y}\in [0, 2]$, but the IR fixed point (line)
belongs to phase III and occurs only for $\t{Y}\in [1,2]$. The positions of the fixed points are  given in Table \ref{tab:fps}.
\begin{table}[htb]
\begin{center}
\begin{tabular}{|c|c|c|}
\hline
 Fixed point & $\t{v}_1$ & $\t{v}_2$ \cr 
\hline\hline
 Gaussian & $0$& $ 0$ \cr
  Wilson-Fisher & $\frac{3}{13}(1-\t{Y})$ & $\frac{80\pi^2}{169}(1-\t{Y})^2$ \cr
  IR & $1-\t{Y}$&$0$\cr
\hline
\end{tabular}
\end{center}
\caption{\label{tab:fps} The positions of the fixed points 
 for given values of the higher-derivative coupling $\t{Y}$.
}
\end{table}
One can see in the top-right subfigure  in  Fig. \ref{fig:phase_structure}
that both the Gaussian and  Wilson-Fisher fixed points lie on the phase boundary III-I and act for the RG trajectories as  cross-over points. The flow of the RG trajectories in phase I is qualitatively the same independently of the value of $\t{Y}$ in the interval $0<\t{Y}\le 2$. In the slice 
$(\t{v}_1, \t{v}_2)$ for $1<\t{Y}\le 2$ the trajectories in phase III run into the IR fixed point (line), but their evaluation becomes numerically unstable in the close neighbourhood of the fixed point.  The phase boundary III-II lies in the plane $\t{Y}=1$.
 Finally in slices  $(\t{v}_1, \t{Y})$ for any constant $\t{v}_2$ (subfigures at the bottom in  Fig. \ref{fig:phase_structure}) one can see all three phases and the triple point. In the 3-dimensional parameter space there is a triple line,
the line of intersection of the phase boundaries III-II and III-I. 
The  detailed study of the IR scaling laws enables one to identify the symmetry properties of the various phases. This is given   in the following
subsections.

\subsection{Phase I}
\begin{figure}[htb]
\centerline{\psfig{file=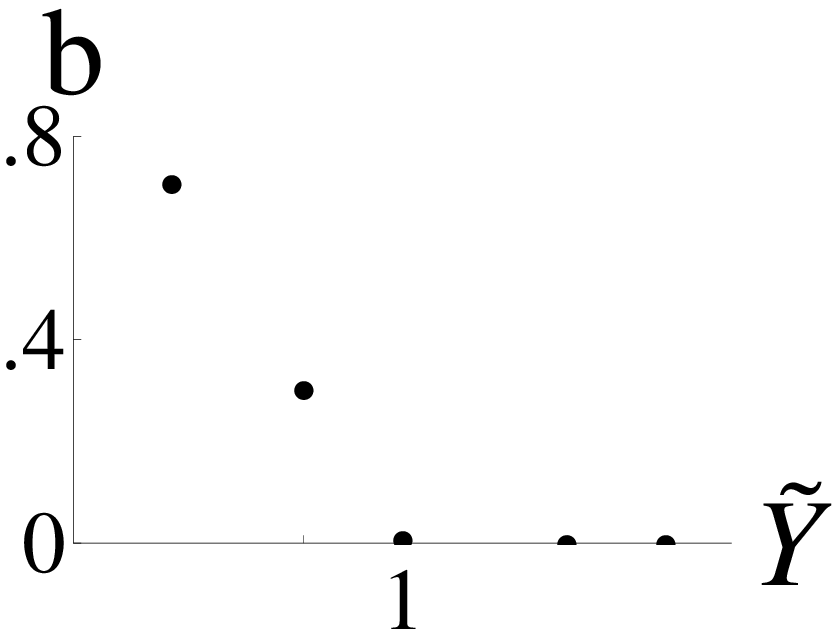,height=3.52cm,width=4.04cm,angle=0}
\psfig{file=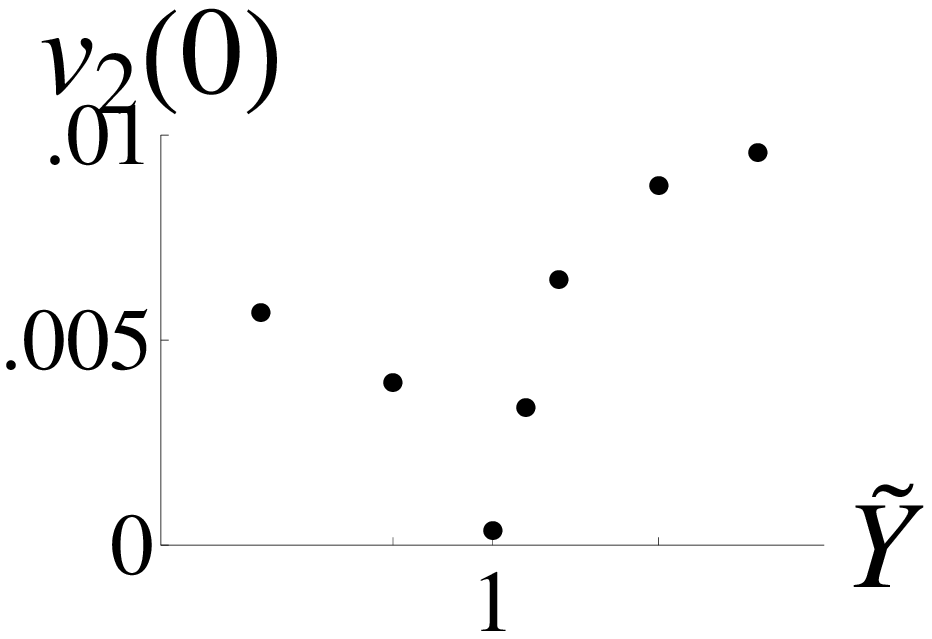,height=3.52cm,width=4.04cm,angle=0}}
\caption{\label{fig:gh_ph1} The parameters $b$ in Eq. \eq{phase1v1} (to the left) and $v_2(0)$ (to the right) vs. the higher-derivative coupling $\t{Y}$
at the `distance' $t=\t{v}_1(\Lambda)-\t{v}_u=0.001$ from the  boundary of phase I. 
}
\end{figure}

 Phase I is the symmetric phase of the  model.
 Trajectories belonging to phase I are those along which the WH RG equation \eq{WHRGeq} does not acquire any singularity. The RG flows of the individual dimensionful couplings are qualitatively the same in phase I, regardless of the value of $\t{Y}$. They increase strictly monotonically with decreasing scale $k$ in a rather short UV scaling region $\sim 0.3 < k\le \Lambda=1$ and then tend asymptotically  to  certain constant values $v_n(0)$  in the IR regime. Therefore the dimensionful effective potential is convex, but very much sensitive to the bare potential. The IR limiting values of the dimensionful couplings $v_1(0)$ and $v_2(0)$ have been compared on
 RG trajectories started at various given `distances'  $t=\t{v}_1(\Lambda)-\t{v}_u$ from the phase boundary $\t{v}_u$ (I-II for $0\le \t{Y}<1$ and I-III for $1<\t{Y}\le 2$) for given values of $v_2(\Lambda)=0.01,~0.1$ and several values of $\t{Y}$. The linear relation 
\bea\label{phase1v1}
  v_1(0)&=& a t +b(\t{Y})
\eea
 has been established where the slope $a=1\pm.001$ is independent of $\t{Y}$, whereas the mass squared at the phase boundary  $(t\to 0)$,
\bea
b(\t{Y})&=&(1-\t{Y})b(0)\Theta(1-\t{Y})
\eea
decreases approximately linearly to zero at $\t{Y}=1$ with the increasing higher-derivative coupling  $\t{Y}$ (see Fig. \ref{fig:gh_ph1}) and
 vanishes for $\t{Y}>1$.

For $k\to 0$ the coupling $v_2(k)$ increases by a great amount
with respect to its bare value $v_2(\Lambda)$  near the phase boundary  I-II for $t\to 0$ and $0< \t{Y}\ll 1$, but it accomodates very little loop-corrections near the boundary  I-III
  for $t\to 0$ and $1<\t{Y}\le 2$. In the latter case the behaviour of the coupling $v_2(k)$  is similar to its behaviour in the symmetric phase of the ordinary $O(2)$ model near the boundary with the symmetry broken phase. Far enough from the phase boundary $\t{v}_u$, i.e., at larger  values of $t$, the loop-corrections are suppressed by the large mass squared $v_1(0)$ and the coupling $v_2(k)$ as well as all higher-order couplings $v_{n>2}(k)$ keep essentially their bare values. For $t\to 0$ the IR value $v_2(0)$ shows up a significant dependence
on the higher-derivative coupling $\t{Y}$, it has a minimum at $\t{Y}=1$ with 
$v_2(0)=0$ (Fig. \ref{fig:gh_ph1}).

\subsection{Phase II}
 Phase II occurs for $0<\t{Y}<1$. As we shall argue below, phase II is  a phase with restored symmetry in the IR limit, i.e.,  a periodic structure breaking $O(2)$ symmetry occurs below the singularity scale $k_s$, but it is washed out in the limit $k\to 0$. In this phase $k_s=\Lambda$, so that the RG flow has to be followed up by the TLR procedure started at the UV scale $\Lambda$. It was found that the couplings of the dimensionful blocked potential tend to constant values in the IR limit.
\begin{figure}[htb]
\centerline{\psfig{file=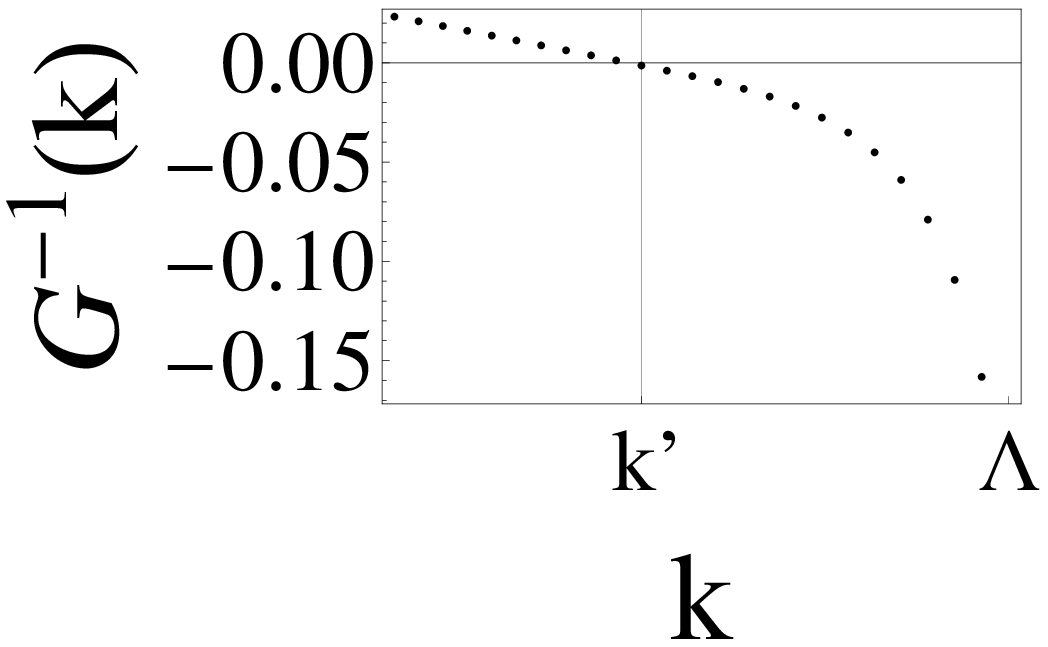,height=3.52cm,width=4.04cm,angle=0}
\psfig{file=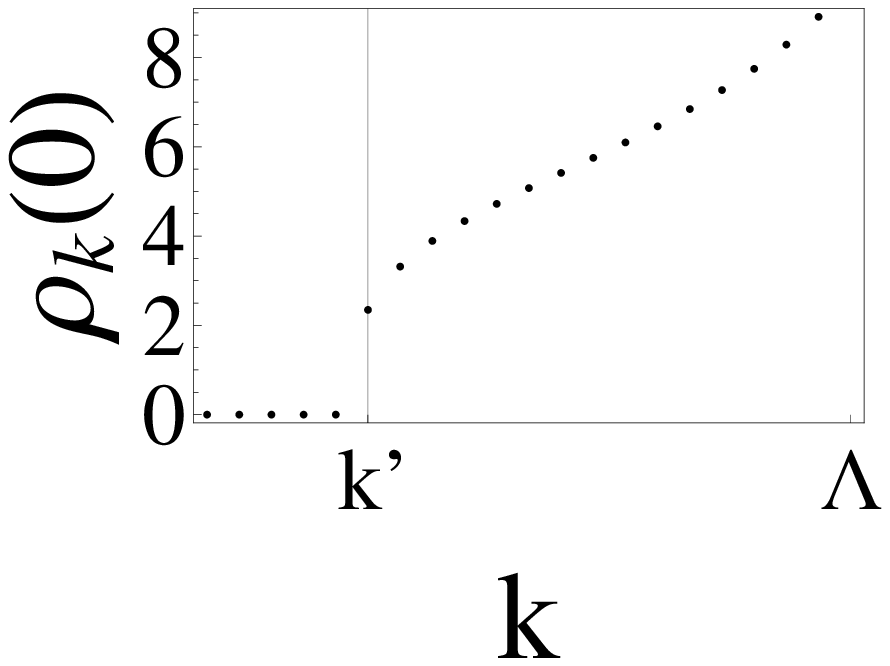,height=3.52cm,width=4.04cm,angle=0}}
\caption{\label{fig:ginv_rho_ph2} The flow of the inverse propagator $G^{-1}(k)$  and that of the amplitude of the spinodal instability $\rho_k$ at vanishing homogeneous background field $\Phi=0$ along the RG trajectory with $\t{Y}=0.7$, $\t{v}_1(\Lambda)=-0.1$, $\t{v}_2(\Lambda)=0.01$ and the step size $\Delta k/k=5\cdot 10^{-5}$. 
}
\end{figure} 
The typical behaviour of the inverse propagator $G^{-1}=(-1+\t{Y})k^2+ v_1(k)$
and that of the amplitude of the spinodal instability $\rho_k(0)$ for vanishing homogeneous background field $\Phi=0$  are shown in Fig. \ref{fig:ginv_rho_ph2}. It can be seen that just below the scale $k_s=\Lambda$ the inverse propagator is negative, its magnitude as well as the amplitude $\rho_k(0)$ decrease till
 the gliding cutoff $k$ reaches  some nonvanishing scale $k'<k_s$. It was found that $\rho_k(0)$ decreases linearly with the scale $k$ in the interval $k'< k<\Lambda$. At the scale $k'$ the propagator vanishes and the amplitude of the spinodal instability  jumps to zero suddenly. This means that below the scale $k'$ no  tree-level renormalization occurs any more. The flow of the amplitude $\rho_k$ of the spinodal instability is qualitatively  just the same as we have found it previously in our paper \cite{Peli2016} in phase II. Namely the periodic configuration is developed below the scale $k_s$ but it is washed out at some nonvanishing scale $k'$.

\begin{figure}[htb]
\centerline{\psfig{file=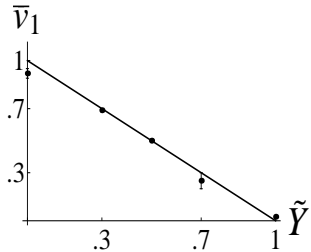,height=3.52cm,width=4.04cm,angle=0}}
\caption{\label{fig:bv10tY} 
The dimensionful mass squared $\b{v}_1(0)$ vs. the higher-derivative coupling $\t{Y}$ in phase II.
}
\end{figure}

 Moreover, it has been observed that for any given value of the higher-derivative coupling $\t{Y}$ the effective potential  is quasi-universal  in the sense that it does not depend on at which point $(v_1(\Lambda),~v_2(\Lambda))$ the RG trajectories have been started. Therefore we have determined the mean values $\overline{v_1(0)}$ and $\overline{v_2(0)}$ of the couplings $v_1(0)$ and $v_2(0)$ with their variances via averaging them over all evaluated RG trajectories belonging to a given value of the coupling  $\t{Y}$.  Table \ref{ty_summtable} shows that the {\em dimensionful} mass squared decreases with increasing values of $\t{Y}$ linearly as
\bea
   \overline{v_1(0)}(\t{Y}) &=& \lbrack \overline{v_1(0)}\rbrack_{\t{Y}\to 0} (1-\t{Y})
\eea
(c.f. Fig.\ref{fig:bv10tY}), while the coupling of the quartic term vanishes. Similarly, all the higher-order couplings $\t{v}_{n>2}(0)$ vanish. One should remind that the theory in the limit $\t{Y}\to 0$ is not  bounded  energetically
from below. 
\begin{table}[htb]
\begin{center}
\begin{tabular}{|r|c|r|}
\hline
 $\t{Y}$ & $\overline{v_1(0)}\pm \Delta v_1(0)$ & $\overline{v_2(0)}\pm\Delta v_2(0)$ \cr 
\hline\hline
  $.0$& $.92\pm .03$& $ -.016 \pm .036$ \cr
  $.3$ & $.69\pm .01$& $-.010\pm .016$ \cr
  $.5$ & $.50\pm .01$&$-.007\pm .016$\cr
  $.7$& $.25\pm .05$ & $.002 \pm .050$ \cr
  $1.0$& $.025\pm .007$ & $-.016 \pm .018$ \cr
\hline
\end{tabular}
\end{center}
\caption{\label{ty_summtable} 
Mean IR values of the dimensionful couplings of the quadratic and quartic terms of the effective potential with their errors in phase II for various values of the higher-derivative coupling $\t{Y}$.
}
\end{table}

It might happen that the loop corrections would become significant for scales $k<k'$ again. Therefore we used the values of the couplings $\t{v}_n(k')$ ($1\le n\le 10$) obtained by the TLR procedure as initial conditions for solving the WH RG equation for $k<k'$. 
It has been established that the loop-corrections cause  less than $0.1$  per cent change in the value of $v_1(k')$ and $\sim 30$ per cent change in $v_2(k')$
on any particular RG trajectory. Since we have argued above that the nonvanishing values of $v_n(k')$ for $n\ge 2$ is due to numerical inaccuracies, we have to conclude that  our TLR result obtained at the scale $k'$ is stable against 
further loop-corrections in the region $0\le k <k'$.

\subsection{Phase III}
\begin{figure}[htb]
\centerline{\psfig{file=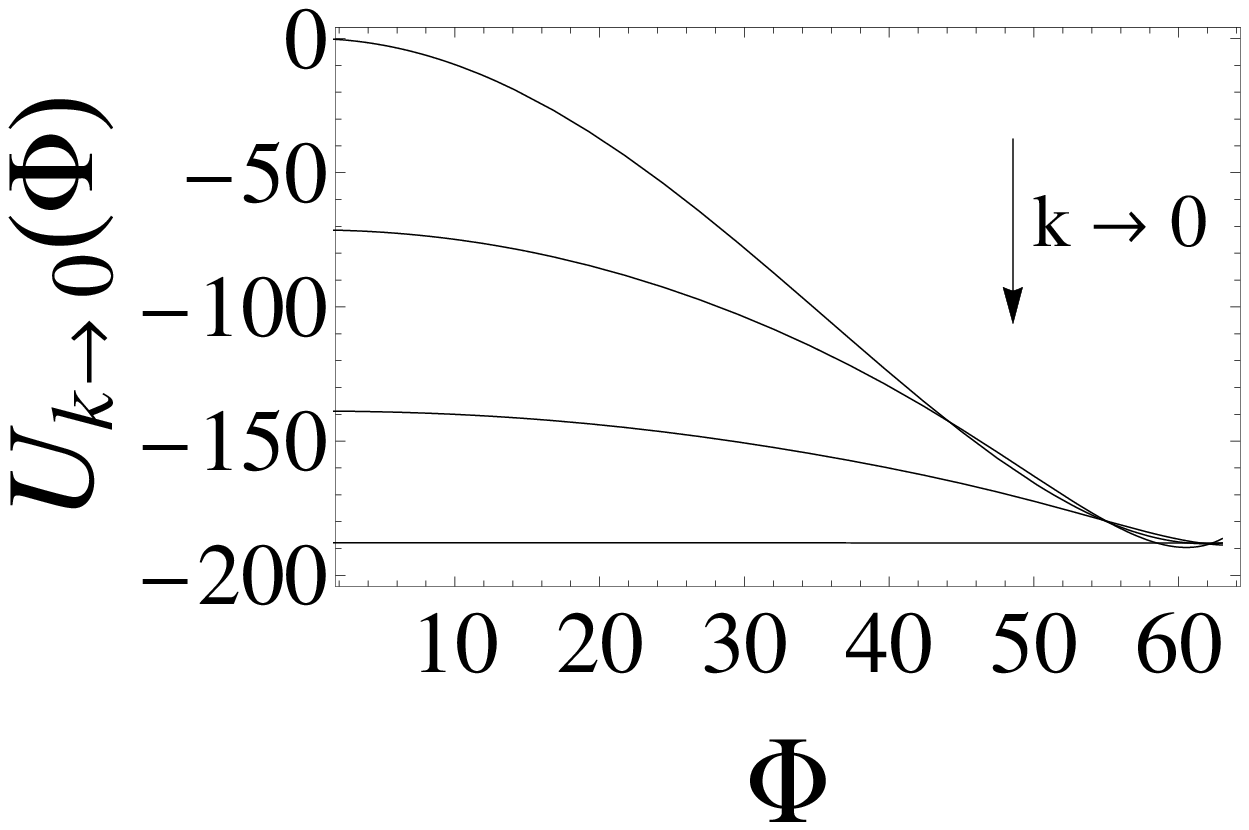,height=3.52cm,width=4.04cm,angle=0}
\psfig{file=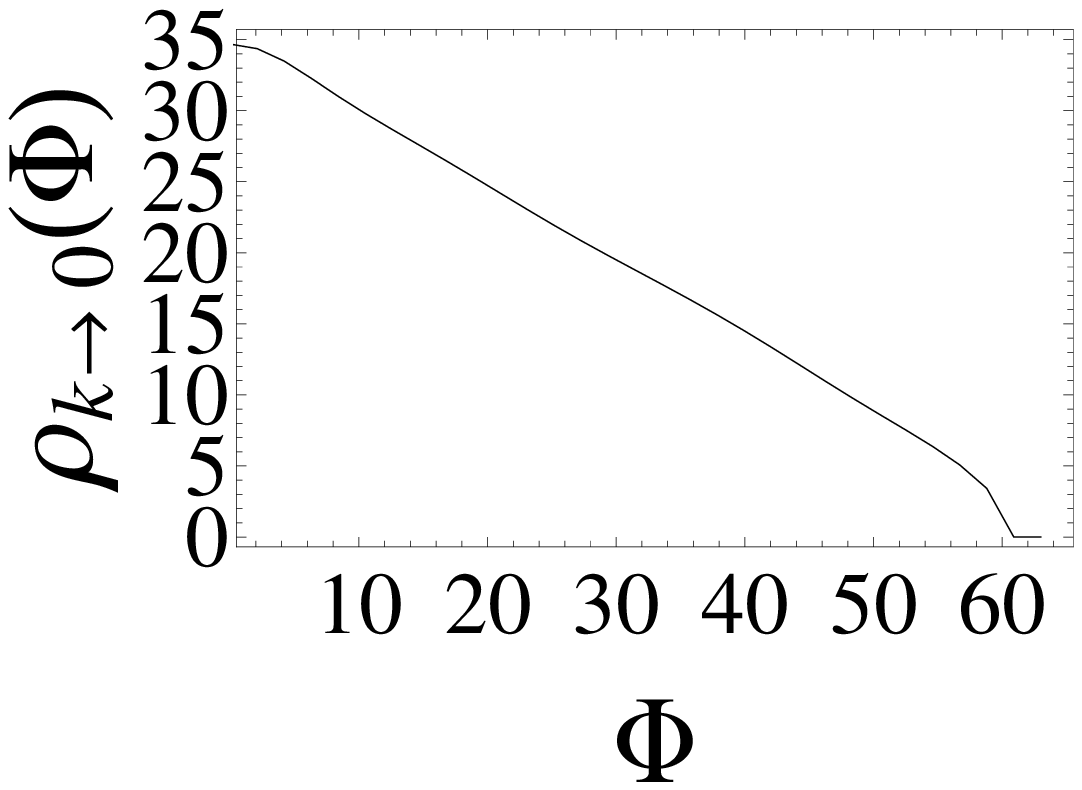,height=3.52cm,width=4.04cm,angle=0}
}
\caption{\label{yt_phase2b} The dimensionless blocked potential $\t{U}_k(\Phi)$
 (to the left) and 
the amplitude $\rho_0(\Phi)$ of the spinodal instability (to the right) vs. the homogeneous background field $\Phi$ for $\t{Y}=1.5$ in phase III.
}
\end{figure}

 Phase III occurs for $\t{Y}>1$ and consists of two regions in the
 parameter plane $(\t{v}_1,\t{v}_2)$ specified by the singularity scale
$k_s=\Lambda$  in the region with $-1\le\t{v}_1\le -1+\t{Y}$ and $k_s=k_c<\Lambda$ for $-1+\t{Y}<\t{v}_1<\t{v}_u$, where $\t{v}_u$ is the phase boundary III-I.   It has been established numerically that phase III is characterized by spontaneous breaking of $O(2)$ symmetry and a quasi-universal {\em dimensionless} effective potential
\bea\label{dlesspot} 
\t{U}_{k\to 0}(\t{\Phi})&=&-\frac{1}{2}(-1+\t{Y})\t{\Phi}^2,
\eea
 providing the Maxwell-cut like universal dimensionful effective potential (see Fig.\ref{yt_phase2b} and the numerical value of $\t{v}_1(0)$ in Table \ref{symbreak_compare} which should be compared with its theoretical value $1-\t{Y}$). 
The dimensionless effective potential \eq{dlesspot} of the form of a downsided parabola with curvature $1-\t{Y}<0$ is the generalization of that
with curvature $-1$
obtained in the symmetry breaking phase of the ordinary $O(2)$ model without higher-derivative terms. The latter case is recovered as a limiting one for $\t{Y}=2$. The presence of the higher-derivative coupling $\t{Y}>1$ result in decreasing the magnitude of the curvature. Like in the case of the ordinary $O(2)$ model, it has been found that the amplitude of the spinodal instability
survives the IR limit and depends linearly on the homogeneous background field
$\Phi$,
\bea\label{rhokphi} 
\rho_{k\to 0}(\Phi)&=&\beta(-\Phi+\Phi_c(0)).
\eea
The values of the coefficient $\beta$ obtained numerically for various values of $\t{Y}$ are compared in  Table \ref{symbreak_compare}. These values do not show up any dependence on the higher-derivative coupling $\t{Y}$ and yield the mean value $\bar{\beta}=-.53\pm .01$. Based on this result and  the assumption that the limit $\t{Y}\to 2$ were continuous  one is inclined to suggest that the exact value is $\beta=1/2$, but our TLR procedure has some systematic error. 

\begin{figure}[htb]
\centerline{\psfig{file=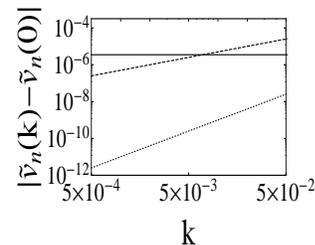,height=3.52cm,width=4.04cm,angle=0}}
\caption{\label{yt_couplings} Scaling of the dimensionless couplings $\t{v}_1(k)$ (full line), $\t{v}_2(k)$ (dashed line) and $\t{v}_3(k)$ (dotted line) for $\t{Y}=1.5$ in the symmetry broken phase III.
}
\end{figure}

 We also  determined numerically the scaling of the dimensionless couplings in the deep IR region (see Fig. \ref{yt_couplings}).
There occurs a clearcut IR scaling region in which the couplings $\t{v}_n(k)$ with $n\ge 2$ scale down to zero according to the power law $\t{v}_{n\ge 2}\sim k^{\alpha_n}$, while  $\t{v}_1(k)-\t{v}_0(0)\sim k^{\alpha_1}$ remains essentially zero in the same region.  
The numerical values of the scaling exponents $\alpha_n$  turned out to be universal, as shown in   Table \ref{symbreak_compare}. This means that all the dimensionful couplings reach their constant IR values with the power law $v_n(k)-v_n(0)\sim k^2$.
\begin{table}[htb]
\begin{center}
\begin{tabular}{|c|c|c|c|c|c|c|c|c|}
\hline
 $\t{Y}$ & $\t{v}_1(0)$ & $\t{v}_2(0)$  &$\alpha_1$ & $\alpha_2$ & $\alpha_3$ & $\alpha_4$ & $\beta$\\ \hline
\hline
 $1.3$  &  $-.281$ & $<10^{-5}$  & $0$ & $1$  & $2$ & $3$ & $.534$\\ 
\hline
 $1.5$  &  $-.469$ & $<10^{-5}$ &  $0$ & $1$  & $2$ & $3$ & $.531$ \\ 
\hline
 $1.8$  &  $-.75$ &  $<10^{-5}$  &  $0$ & $1$ & $2$ & $3$ & $.521$ \\ 
\hline
 $2$  &  $-.94$ &  $<10^{-5}$  & $0$ & $1$ & $2$ & $3$ & $.531$\\ 
\hline
\end{tabular}
\end{center}
\caption{\label{symbreak_compare} The IR limiting values of the first two 
couplings of the dimensionless potential, the coefficient $\beta$ of the amplitude in Eq. \eq{rhokphi}, and the first few scaling exponents $\alpha_n$ 
obtained by TLR for phase III.}
\end{table}

\begin{figure}[thb]
\centerline{\psfig{file=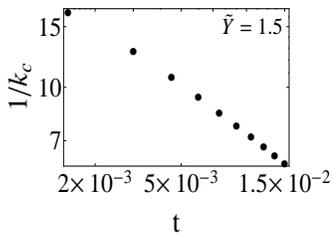,height=3.52cm,width=4.4cm,angle=0}
}
\caption{\label{fig:critical_exp} Scaling of the correlation length $\xi\sim 1/k_c$ 
with the reduced temperature $t=\t{v}_u-\t{v}_{1}(\Lambda)$ (on a log-log plot) at the boundary of phases I and III  for $\t{Y}=1.5$ and $\t{v}_2(\Lambda)=0.1$.
}
\end{figure}

\subsection{On the phase transitions}

In thermodynamics phase transitions accompanied by a finite jump of the free energy, i.e., the presence of  a latent heat are called of first order, while those with continuous free energy and singularities in the derivatives of the free energy are continuous. As to the transition from phase III to phase I in our case, there is a rather straighforward way to decide that the transition III-I is continuous. Namely, one  determines the behaviour of the correlation length $\xi\sim 1/k_c$ approaching the boundary of phases I and III from the side of phase III. This approach is applicable only at the phase boundary III-I, because the singularity scale
$k_c$ can be detected by solving the WH RG equation \eq{WHRGeq}, while this scale lies above the UV cutoff $\Lambda$ for phase II, therefore we cannot make such calculations at the phase boundaries II-I and II-III. The reduced temperature is identified
as $t=\t{v}_u- \t{v}_{1}(\Lambda)$, i.e., the `distance' of the starting point of the RG trajectories from the phase boundary $\t{v}_u$. In order to determine the dependence of the correlation length $\xi$ on the reduced temperature $t$
 we have solved the WH RG equation \eq{WHRGeq}  for various initial conditions    $\t{v}_{1i}(\Lambda)=\t{v}_u-i\cdot 10^{-4} $ $(i=1,2,\ldots, 500)$ for each values of $\t{v}_2(\Lambda)=0.01,~0.1$ and  $\t{Y}=1.2,~1.5,~2.0$. 
It has been established that the correlation length has a power law behaviour 
\bea\label{corpowlaw}
   \xi\sim 1/k_c&\sim& t^{-\nu},
\eea
near the phase boundary for any fixed values of the coupling $\t{Y}$ (see Fig. \ref{fig:critical_exp}). This signals that the phase transition III $\to$ I is continuous, just like the transition in the ordinary $O(2)$ model. The critical exponent $\nu$ seems to be insensitive to the bare parameters $\t{Y}$ and $\t{v}_2(\Lambda)$, its mean value is $\bar{\nu}=.46\pm .03$. 
 The $\phi^4$ model can be considered as the textbook example of the RG technique. Therefore, it is widely investigated in various dimensions and in various levels of truncations \cite{Tetradis1994,Liao,Litim2001,Canet2003,ZJ,Morris1997,Litim2011,Ber2005,Mati2016}. Our result is close to the best value $\nu=0.63$ obtained in the 3-dimensional case.

\begin{figure}[thb]
\centerline{\psfig{file=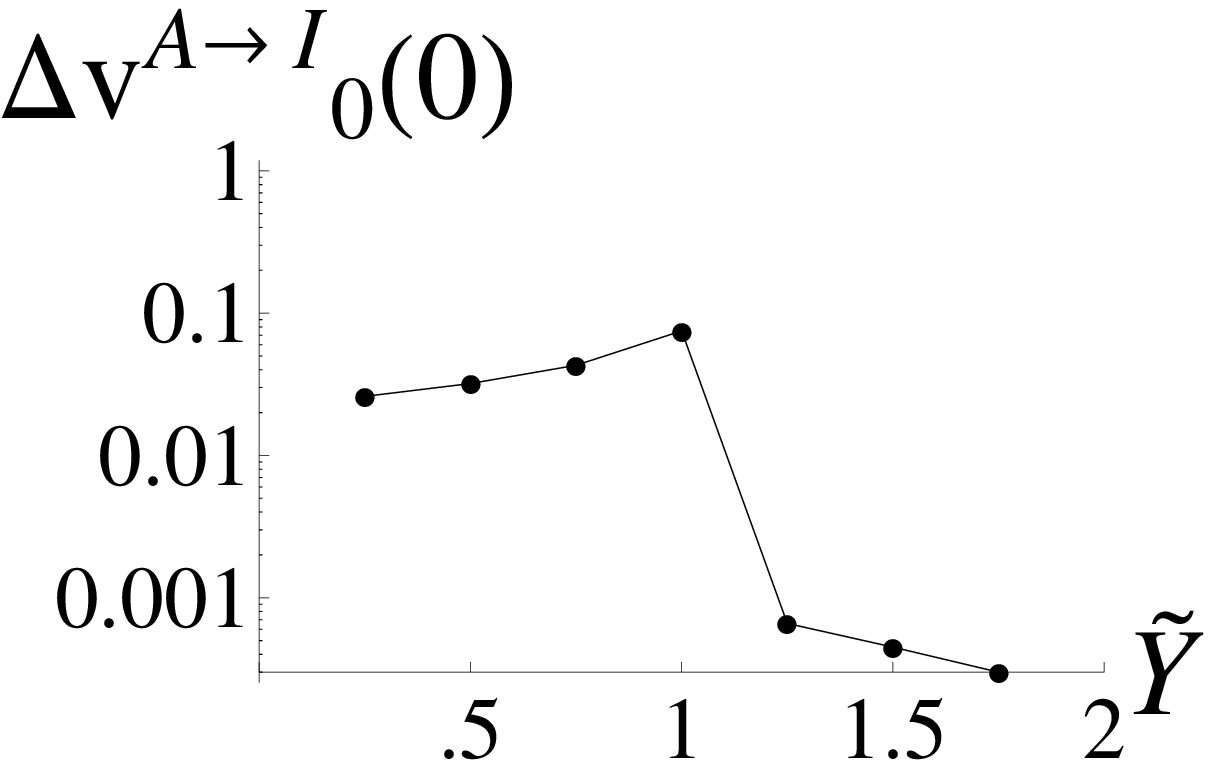,height=3.52cm,width=4.4cm,angle=0}
\psfig{file=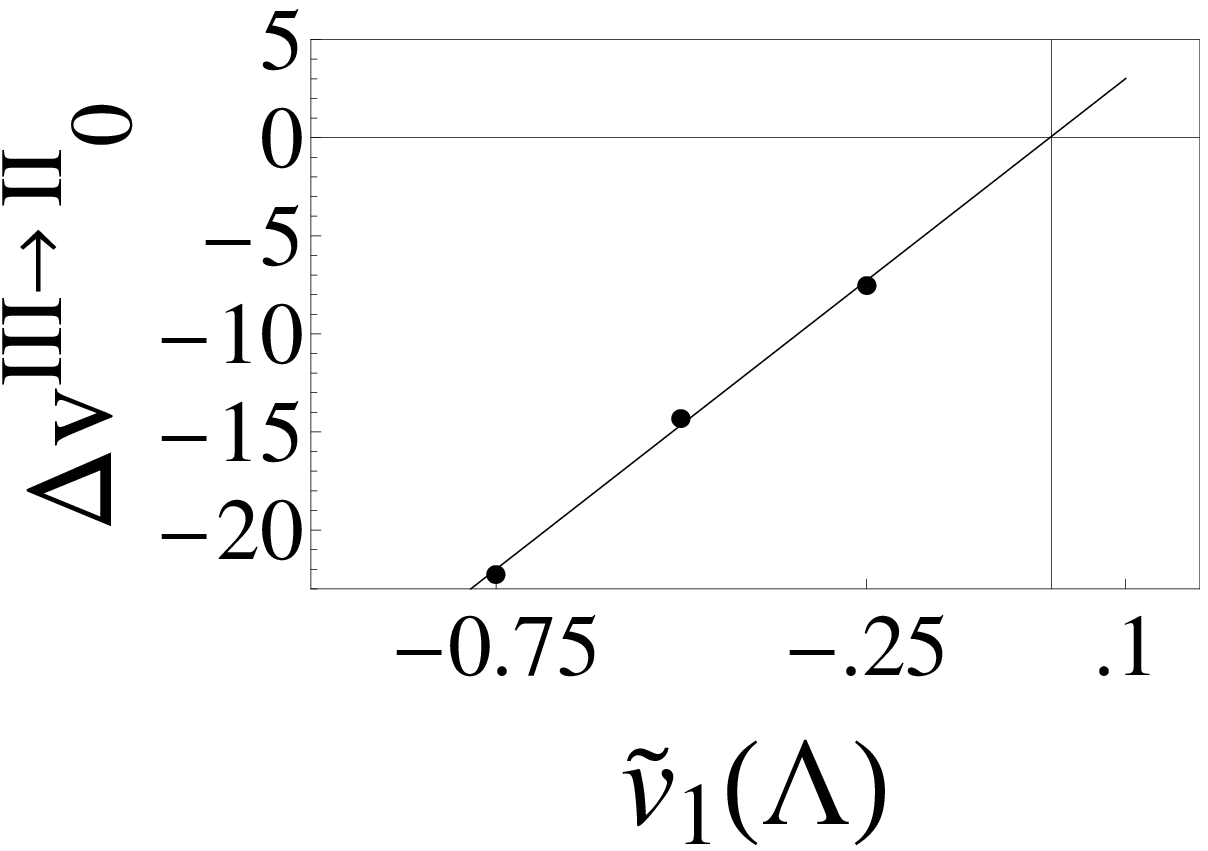,height=3.52cm,width=4.4cm,angle=0}
}
\caption{\label{fig:trans} The jump of the `free-energy density' $\Delta v_0^{A\to B}$ for $B=$I, $A=$II ($0<\t{Y}<1$) and $A=$III ($1<\t{Y}<2$) (to the left),
and for $A=$III, $B=$II (to the right).   
}
\end{figure}

There is however another way to study the continuity of the phase transition.
Namely, we can determine directly the jump of the free-energy density or latent heat per unit volume, more precisely
the jump of minimum of the effective potential going from one phase to another across the phase boundary.  For that purpose we have to determine the IR limit
of the constant term $v_0^A(0)$ of the effective potential in the various phases
$A=$I,~II,~III at both sides of the phase boundary and compare them. For the comparison we have to consider RG trajectories on which the bare potential has the same minimum value. Otherwise, this can be put as the correction of the IR values $v_0^A(0)$
by the minimum value of the bare potential $(U_\Lambda^A)_{min}$, i.e., by the replacement $v_0^A(0)\longrightarrow (v_0^A)_{corr}=v_0^A(0)-(U_\Lambda^A)_{min}$.
The transition from phase $B$ to phase $A$ is then accompanied by the jump
of the potential (Euclidean action per volume) $\Delta v_0^{A\to B}= (v_0^B)_{corr} - (v_0^A)_{corr}$. The nonvanishing or vanishing value of $ \Delta v_0^{A\to B}$
signals that the transition is of first order or continuous, respectively.
In our settings $(U_\Lambda^A)_{min}$ is nonvanishing only for RG trajectories
belonging to bare potential of double-well form (those starting close to 
the phase boundaries III-I and III-II in phase III, and close to the phase boundary II-III in phase II). For the numerical determination of $\Delta v_0^{\rm{II}\to {\rm{I}}}$
we have chosen RG trajectories which start at the `distance' $t=0.001$ from the phase boundary. In the case of the evaluation of $\Delta v_0^{\rm{III}\to \rm{I}}$, we
considered RG trajectories with the values of $v_1(\Lambda)$ increased in steps
$t=0.001$ crossing the phase boundary. Finally, $\Delta v_0^{\rm{III}\to \rm{II}}$ has been determined from the comparison of RG trajectories for $\t{Y}=1.1$ and $0.9$
and various values of $\t{v}_1(\Lambda)$. All calculations were made for $\t{v}_2(\Lambda)=0.01$.
 The results are shown in Fig.
\ref{fig:trans}. In the plot on the left we see that there is a jump of the free-energy density of 2 orders of magnitude larger for $0<\t{Y}<1$ than for $1<\t{Y}<2$. Together with our previous finding on the base of the study of the correlation length this enables one to conclude that the phase transitions
III$\to$ I and II$\to$ I are continuous and of first order, respectively.
 Similarly, the plot to the right in Fig. \ref{fig:trans} shows that the phase transition III$\to $II is of first order with a latent heat per unit volume decreasing to zero when the triple point is approached. 
  
\section{Conclusions}

The phase structure of  the 3-dimensional Euclidean  $O(2)$ symmetric ghost scalar field model has been investigated in the framework of the Wegner and Houghton's (WH) renormalization group (RG), including the higher-derivative term $- \hf \int_x Y\phi\Box \phi$ into the action and keeping the dimensionless coupling $\t{Y}$ constant. The RG flow with decreasing gliding cutoff $k$ has been determined numerically by solving the WH RG equation. When the right-hand side of the WH equation develops a singularity at some scale $k_c\not=0$, the flow has been followed further by means of the tree-level renormalization (TLR) procedure. It has been shown that the model exhibits three phases and a triple line.
 The symmetric phase (phase I) is present for any values $\t{Y}>0$ and shows similar features to the symmetric phase of the ordinary $O(2)$ model.  Phase II is present, when $0\leq\t{Y}\leq 1$. The RG flow of the trajectories belonging to phase II can only be determined by  the TLR procedure
on all scales below the UV cutoff  $\Lambda$. The dimensionful effective potential in phase II is quasi-universal, it depends on the value of the higher-derivative coupling $\t{Y}$, but is independent of the other bare couplings. Just below the scale $\Lambda$ there occurs a periodic spinodal instability that breaks $O(2)$ symmetry as well as rotational and translational symmetries in the external space, however, that intermediate symmetry breaking is washed out  in the IR limit.  Phase II has no analogue in the ordinary $O(2)$ model. It is of the same properties found in \cite{Peli2016} and its existence is based upon the ghost-condensation mechanism available in the model with $Z<0$ and $\t{Y}>0$. 
 Phase III occurs for $1<\t{Y}\leq 2$. It separates into two regions, region IIIA and IIIB, where TLR has to be used below the  UV cutoff  $\Lambda$ and the singularity scale  $k_c$, respectively, however, both regions  IIIA and IIIB have the same deep IR behaviour.  In phase III the dimensionful effective potential is universal, it exhibits the Maxwell cut which is accompanied  with the nonvanishing amplitude of the periodic spinodal instability for scales $k\to 0$. Therefore phase III is the one in  which spontaneous symmetry breaking occurs, just like in the symmetry breaking phase of the ordinary $O(2)$ model. The phase boundaries III-I and III-II
intersect in a triple line.

It has been studied the continuity of the various phase transitions
by means of the differences of the minimum values of the effective potentials in the various phases
and found that the phase transitions II$ \to$ I and III$\to$ II are of the first order accompanied with a nonvanishing latent heat per unit volume, whereas
the transition III$\to$ I is continuous. The latter has been also supported by the scaling behaviour of the correlation length in phase III close to the phase boundary III-I.

In the framework of the WH RG restricted to the local potential approximation (LPA), 
the phase structure of the model turned out to be  more rich when the dimensionless higher-derivative coupling $\t{Y}$ is kept constant during the RG flow than
in the case when the dimensionful coupling $Y$ is kept constant, as we did in our previous work \cite{Peli2016}. Therefore, it remains an open question whether the model exhibits two or three phases.
 The ambiguity of keeping constant either the dimensionful or the dimensionless higher-derivative coupling  is an essential feature of the local potential approximation (LPA) and is unavoidable in the WH RG approach \cite{Pol2003}.
Our work demonstrates that such an ambiguity may affect the physical results significantly when higher-derivative terms are included into the model. No similar ambiguity should occur if one goes  beyond the LPA in the gradient expansion
using any RG framework being appropriate for it, e.g., the effective average action approach \cite{Tet1992} .

\section*{Acknowledgements}
S. Nagy acknowledges financial support from a J\'anos Bolyai Grant of the
Hungarian Academy of Sciences, the Hungarian National Research Fund OTKA
(K112233).

\appendix

\end{document}